\documentclass[aps,pre,preprint,groupedaddress]{revtex4-1}
\usepackage{epsfig}
\usepackage{rotating}
\usepackage{color}
\usepackage[normalem]{ulem}

\begin{document}

\title{Beyond Zipf's Law: The Lavalette Rank Function and Its Properties }

\author{Oscar Fontanelli$^1$, Pedro Miramontes$^{1,2}$, Yaning Yang$^3$, Germinal Cocho$^4$, Wentian Li$^5$}
\affiliation{1. Departamento de Matem\'{a}ticas, Facultad de Ciencias,
Universidad Nacional Aut\'{o}noma de M\'{e}xico,
M\'{e}xico
DF, M\'{e}xico; \\
2. Bioinformatics Group and Interdisciplinary Center for Bioinformatics,
University of Leipzig,
Leipzig, Germany;
3. Department of Statistics and Finance, University of Science and
Technology of China, Hefei, Anhui, China; \\
4.  Instituto de F\'{i}sica, Universidad Nacional Aut\'{o}noma
de M\'{e}xico, M\'{e}xico
DF, M\'{e}xico;
5. The Robert S. Boas Center for Genomics and Human Genetics,
The Feinstein Institute for Medical Research,
Northwell Health, Manhasset, NY
USA.
}

\date{\today}

\begin{abstract}

Although Zipf's law is widespread in natural and social data, one often encounters situations
where one or both ends of the ranked data deviate from the power-law function. Previously we
proposed the Beta rank function to improve the fitting of data which does not follow a perfect
Zipf's law. Here we show that when the two parameters in the Beta rank function have the same value,
the Lavalette rank function, the probability density function can be derived analytically. We also
show both computationally and analytically that Lavalette distribution is approximately equal,
though not identical, to the lognormal distribution. We illustrate the utility of Lavalette
rank function in several datasets. We also address three analysis issues on the statistical
testing of Lavalette fitting function, comparison between Zipf's law and lognormal distribution
through Lavalette function, and comparison between lognormal distribution and Lavalette
distribution.

\end{abstract}


\maketitle

\section*{Introduction}
It is said that a certain quantity follows a power law if the probability of observing  it varies inversely as a power of this quantity. Power laws in data collected from natural or social phenomena are well documented \cite{clauset}.  For instance, the asymptotic occurrence of power laws in critical phenomena and statistical physics has been widely studied \cite{sornette}.  In the same way, power law tails have been reported in the distribution of word frequency \cite{zipf}, city sizes \cite{gabaix}, fluctuations in financial market indexes \cite{gopikrishnan}, firm sizes in the U.S \cite{axtell}, scientific citations \cite{petersen,petersen13} and there are many other examples.  There are two common approaches in displaying a power law distribution: the histogram, which approximates the probability density function (pdf), and the rank-frequency plot, best known by the Zipf's law for usage of words in human languages \cite{zipf,wli_glo}.

Empirical data often exhibit good power-law distribution within a limited range, whereas one or both ends of the distribution may deviate from the ideal power law \cite{stumpf}. It is a well known fact that any finite size system, that is well described by a power law, deviates from this behaviour
due to finite size effects \cite{laherre}. In these systems, the power law ceases to hold in a certain region, where effects due to the finiteness of the system dominate the behaviour (for example, 
finite sample size or finite available energy). Therefore, it is natural to see deviations from power laws at the tails. However, the question remains of whether deviations are merely explained by finite size effects or if they call for a modification in the whole body of the distribution. This paper explores the second possibility. Modifying a power law by changing the functional form potentially may fit the systematic deviation. Previously, we proposed a rank-frequency function, inspired by the Beta density function \cite{beta}, called Beta-like function \cite{beta-joi}, or Discrete Generalized Beta Distribution (DGBD) \cite{beta-plos1}, or Cocho rank function \cite{beta-jql1}. The DGBD

\begin{equation}
\label{eq-beta}
x_{[r]}=C \frac{(N+1-r)^b}{r^a }
\end{equation}

($x$: quantity of interest, $r$: rank, $N$: 
the maximum rank), contains the fitting parameters $a$ and $b$ and the normalization factor $C$.  We previously proposed that the parameter $a$ is associated with the behaviour which leads to the power law, whereas $b$ is associated with the fluctuation in noise \cite{beta-plos1}. An example of the former is the inertial range in turbulence where energy is transferred between different length scales with the same rate, while an example of the latter is the dissipative range in turbulence \cite{beta-plos1}. Another example is in a conflicting dynamics called expansion-modification systems \cite{expmod}, 
where $a > b$  when expansion dominates mutation and $b > a$ when mutation dominates \cite{roberto}.
Eq.(\ref{eq-beta}) modifies the power law rank function $1/r^a$ by a power of the reverse-rank $r_2= N+1-r$, and it converges to power law when $b=0$. DGBD often surpasses other two-parameter functions in fitting real data \cite{beta-jql1,beta-entropy,petersen}, and achieved various degree of success in other applications \cite{beta-joi,beta-mus,beta-plos1,petersen,beta-pra,beta-com,beta-jql2,beta-bmcbi,beta-ausloos}.

It is a well known fact that a quantity that follows a power law in the rank-frequency \-re\-pre\-sen\-ta\-tion has a Pareto distribution \cite{newman}.  The widespread application of the DGBD raises the issue of whether it is the result from a well known pdf, such as the normal/Gaussian distribution.
In this work, we show that for a special case of the DGBD, the Lavalette rank function where
$a=b$ \cite{lava,pope97,lava-pope,lava07,lava-volo}, the corresponding pdf can be derived analytically. The Lavalette rank function is also intrinsically connected, by an approximation,  to the lognormal distribution. We offer both numerical evidence and an analytic proof.

The paper is organized in the following way: first we derive and characterize the pdf associated with the Lavalette rank function, which we call the Lavalette distribution, and show that it is approximately 
equal to the lognormal distribution over a relatively large interval. Next we exhibit applications of the Lavalette distribution to real data, coming from natural and social phenomena, and we discuss a goodness of fit test to prove that this distribution is consistent with the data. Finally, we propose a method for discerning between Lavalette and lognormal distributions and discuss the implications of our findings.

\section*{Results}

The two representations of a distribution, pdf and rank-frequency plot, can be converted from one to the other in these two ways: (i) equating cumulative distribution function (cdf) to reversed normalized rank:
$ r_{[x]}/N \approx  1- \int_{-\infty}^x p(t) dt$; (ii) equating the averaged rank of a value $x$,
$\langle  r_{[x]} \rangle$,  to the $n$ which maximizes the following probability:
$(N-n+1) \left( \begin{array}{c} N \\ n \end{array}\right)
\left( \int_{-\infty}^x p(t)dt \right)^{N-n} \cdot p(x) \cdot
\left( \int_x^{\infty} p(t)dt \right)^{n-1} $ \cite{sornette}. Below, we will only use (i) in deriving a relationship between the pdf and  the rank-frequency representations.

The Lavalette rank function:

\begin{equation}
\label{eq-lavalette}
x_{[r]}= C \left( \frac{N+1-r}{r} \right)^a
\end{equation}

can be converted to

\begin{equation}
\label{eq3}
\frac{r_{[x]}}{N} = \frac{N+1}{N} \frac{1}{1+ \left( \frac{x}{C} \right)^{1/a}}
\approx \frac{1}{1+ \left( \frac{x}{C} \right)^{1/a}},
\end{equation}

with the right-hand-side being 1-cdf. The pdf is then the negative derivative
of Eq.(\ref{eq3}).

\subsection*{The Lavalette Distribution}

A certain quantity $X$ follows a Lavalette rank function if its rank-frequency or rank-size function 
is a DGBD Eq(\ref{eq-beta}) with equal parameters $a=b$ (\cite{lava}). As we saw, the pdf of $X$ is proportional to the negative derivative of the inverse $r_{[x]}$. We say that a random variable $X$ has a Lavalette distribution with  parameters $C$ and $a$ if it has the density

\begin{equation}
\label{eq-lpdf}
p(x)_{lav} = \frac{1}{aC}
 \frac{ x^{1/a-1}}{\left(1+ \left( \frac{x}{C} \right)^{1/a} \right)^2}.
\end{equation}

With the analytic expression of Eq.(\ref{eq-lpdf}), many properties of the Lavalette
distribution can be easily obtained. The $i$-th moment is:

\begin{eqnarray}
E[ x^i ] &=& \int_0^\infty x^i p(x)dx = \frac{C^i}{a} \int_0^\infty
 \frac{ (x_1)^{1/a-1+i}}{\left(1+ (x_1)^{1/a} \right)^2} dx_1
 \nonumber \\
& =& iC^i \int_0^\infty  \frac{(x_1)^{i-1}}{1+(x_1)^{1/a}} dx_1 \nonumber \\
&=& aiC^i \int_0^\infty \frac{(x_2)^{ai-1}}{1+x_2} dx_2 = \frac{ a i C^i \pi   }{ \sin \left( a i \pi \right) }
\end{eqnarray}

\begin{figure}[!h]
\epsfig{file=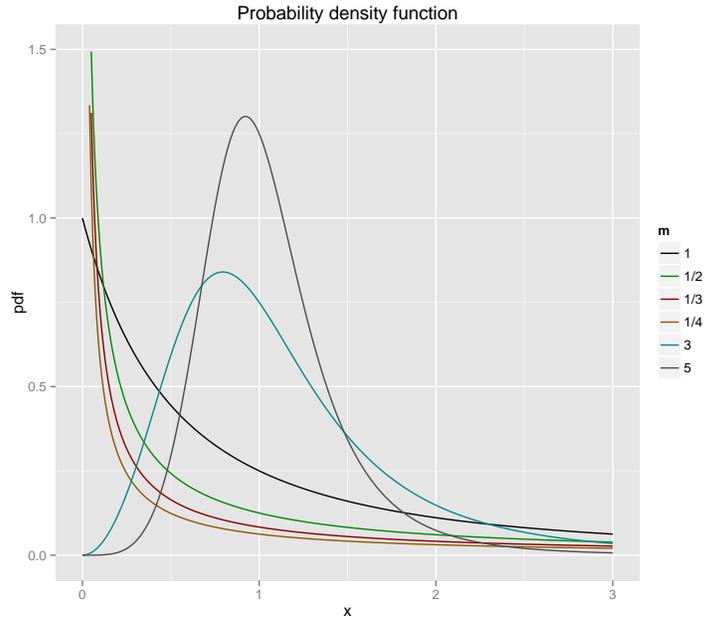,width=0.6\linewidth}
\caption{{\bf Pdf of the Lavalette distribution.}
Some Lavalette probability density functions (Eq.(\ref{eq-lpdf}) with identical
parameter $C=1$ but with a=1/5, 1/3, 1, 2, 3, and 4
($m=1/a=1/b$).}
\label{fig1}
\end{figure}

($x_1= x/C, x_2= x_1^{1/a}$) (if $i < 1/a$) (see, e.g., \cite{integral-table} or
\url{http://en.wikipedia.org/wiki/List\_of\_definite\_integrals}). In particular, the mean of a Lavalette random variable is 

$$
E[x]=  \pi a C \sin(\pi a )^{-1},
$$ 

which is finite if $a < 1$, while its variance is

$$
Var[x]= \pi a C^2 \left( 2 \sin(2a\pi)^{-1} - \pi a \sin( a \pi)^{-2} \right),  
$$

which exists and is finite if $a < 1/2$. However, similar to the discussion of power law distributions,  whether the moments diverge to infinity or do not depends on whether a  lower bound of the functional form is imposed \cite{clauset}. One may re-derive the connection between ranked data and pdf by
$r_{[x]}/N = 1 - \frac{(N-1)}{N} \int_{x_m}^x p(t)pt/ \int_{x_m}^{x_M} p(t)dt$ where $x_m$ and $x_M$ are the minimum and maximum values among $N$ samples.  Fig.\ref{fig1} shows a plot of the Lavalette density for different parameters:  they all have identical $C=1$ but  $a=b= 1/3, 1/5$ (unimodal)
and $a=b=1, 2, 3, 4$ (monotonically decaying).

\subsection*{Resemblance between Lavalette and lognormal distributions}
\label{sec:lavalette-lognormal}

\begin{figure}
\epsfig{file=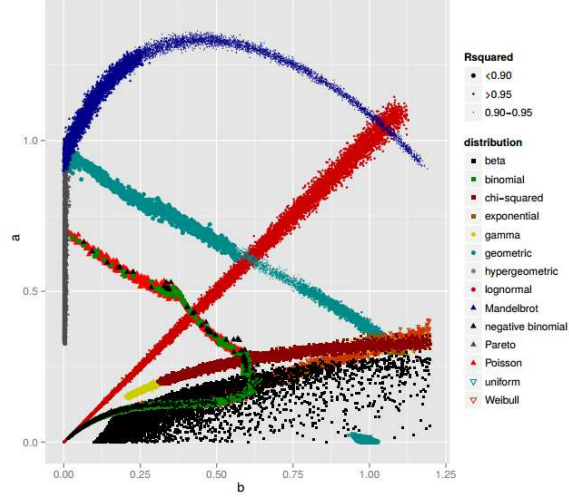,width=0.6\linewidth}
\caption{{\bf DGBD fits for various distributions.}
The estimated $a$ and $b$ parameter values in DGBD (Eq.(\ref{eq-beta})) for data generated by well known distributions. Size of the dots indicate the coefficient of determination R squared.  The dots around the $a=b$ diagonal line are for data generated by the lognormal distribution.}
\label{fig2}
\end{figure}

To examine which well known pdf's share the same property of $a=b$ when fitted to the DGBD rank function, we generated data from 14 distributions (beta, binomial, $\chi^2$, exponential, gamma, geometric,
hypergeometric, lognormal, Mandelbrot, negative binomial, Pareto,
Poisson, uniform and Weibull), and fit the ranked data by the Beta rank function via linear regression
of the logarithmic transformation of Eq.(\ref{eq-beta}). The estimated parameter values for $a$ and $b$ are shown in Fig.\ref{fig2}. Interestingly, the only known pdf which exhibits $a \approx b$ is the lognormal distribution.

We use a novel argument from statistics to explain why the Lavalette and lognormal
distributions may be difficult to distinguish within a certain interval of their domain.
There are two models for probability of a binary variable $y \in (0,1)$:
(i) probit model \cite{probit}: $P = P(y=1) = \Phi (z)$ where $\Phi$ is the cdf of standard
normal distribution; (2) logit model or logistic regression \cite{glm}: $P = 1/(1+e^{-z})$.
The two regression models for binary variable (regressed over an independent variable
$z$) usually lead to similar results \cite{aldrich,agresti}, which can
be written as (after the logistic variable being re-scaled by a factor $\alpha$):

\begin{equation}
\label{eq-logit}
\Phi(z) \approx \frac{1}{1+ e^{-\alpha z}},
\hspace{0.1in} {\mbox or,} \hspace{0.1in}
\frac{ \Phi(z)}{1-\Phi(z)} \approx e^{\alpha z}.
\end{equation}

The $\alpha$ can be $\sqrt{8/\pi} \approx 1.596$ to achieve the
best fit near the midpoint \cite{page},  or $\approx 1.7$ to best fit
the whole range, or $\pi/\sqrt{3} \approx 1.81$ which is the standard
deviation of the variable from the logistic distribution \cite{agresti}.
The standard normal variable can be converted to a lognormal distribution variable $x$:
$z = (\log(x) - \mu)/\sigma$, and re-expressing Eq.(\ref{eq-logit}) in $x$ becomes:

\begin{equation}
e^\mu \left( \frac{\Phi( (\log(x)-\mu)/\sigma}{1-\Phi( (\log(x)-\mu)/\sigma ) }\right)^{\sigma/\alpha} \approx x,
\end{equation}

which we recognize as the Lavalette rank function over variable $x$
($1-\Phi$ is the normalized  rank). This derivation also points out that
the parameter $a=b$ is the standard deviation of the lognormal
distribution divided by $\alpha ( = 1.6 \sim 1.8)$, whereas the log-mean of
the lognormal distribution is related to the scaling parameter by $C=e^\mu$.
Since probit and logistic regression are not the same, we conclude that
the Lavalette and the lognormal distributions cannot be
identical. Indeed, the Lavalette and lognormal distributions have
qualitatively different behaviours at the tails. All moments of the lognormal distribution
exist, while the Lavalette has only finite moments of order $i<1/a$, as we previously discussed. If there is enough data to sample the tail, they cannot be mistaken into one another.\\

\begin{figure}[!h]
\epsfig{file=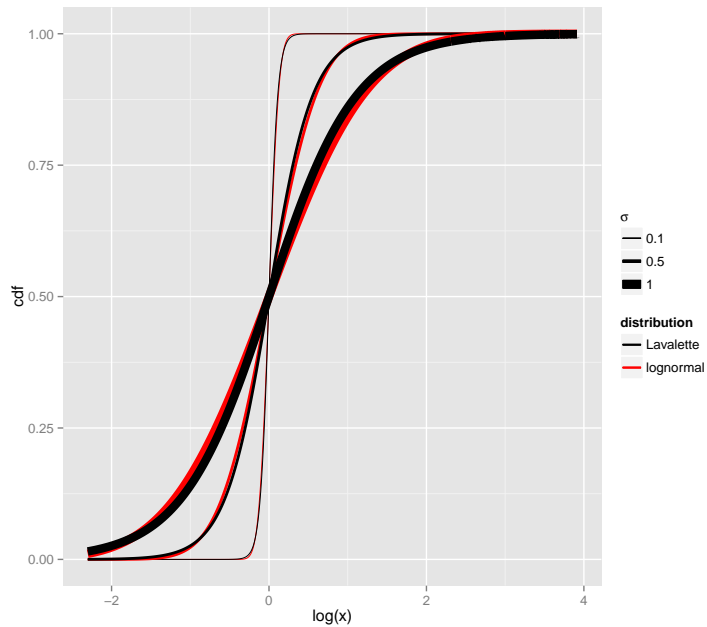,width=0.6\linewidth}
\caption{{\bf Lognormal vs Lavalette cdf.}
Cumulative distribution function for lognormal and Lavalette distributions,
being $\mu=0$ and $\sigma=0.1,\, 0.5$ and $1$ the parameters of the lognormal. The $x$ axis
is in logarithmic scale. We see that over an important interval of the domain, it may be difficult
to distinguish a lognormal from a Lavalette distribution.}
\label{fig3}
\end{figure}

Fig.\ref{fig3} illustrates directly the similarity between the Lavalette and lognormal
distributions. The cdf's of lognormal distribution and the corresponding Lavalette distribution
are plotted at three different parameter values ($\mu=0$ with $\sigma =0.1,\, 0.5$ and $1$ for the lognormal, corresponding $C=e^{\mu},\, a=\sigma\sqrt{3}/\pi$ for the Lavalette).
Besides the difference at the tails (which is not visible from the cdf plot because
the difference along the $y$-axis is very small for extreme values), the two functions
also deviate slightly from each other in the middle range. This deviation is equivalent,
after a transformation, to that between the cdf of standard normal distribution and
logistic function. It has been proposed that a modification of the logistic function,
$1/(1+ exp(-1.5876x - 0.070566 x^3 ))$, is a very good approximation of the
cdf of standard normal distribution \cite{johnson,page}. The small coefficient
of the high-order term is another indication that the cdf of normal and logistic function,
or equivalently, lognormal and Lavalette distributions, are close.

\subsection*{Occurrence and Applications}
\label{sec:appearance}

To illustrate that Lavalette distribution can be applied to real data, we examine \-se\-ve\-ral datasets besides the impact factor and citation data used in \cite{lava,pope97}. We will give examples of population data,  amino acid mutation rates and codon usage data where the Lavalette distribution 
is a good statistical model. The parameters were estimated through linear  regression of the logarithmic transformation of Eq.(\ref{eq-beta}), which in our case gives very similar results to maximum likelihood estimators.  The goodness of fit tests were performed using the Kolmogorov-Smirnov statistic and the $p-$values were estimated through a Monte-Carlo  approach proposed in (\cite{clauset}). As usual, a small $p-$ value leads to reject the hypothesis that the data are well described by a Lavalette distribution. 

The first set of examples is about administrative units of population. Most countries in the world are internally divided into administrative units, which may be called states, provinces, etc. \cite{glaw}. We call these primary administrative units (PAU), which may be in turn subdivided into smaller or second level administrative units (counties, municipalities, etc.) We call these secondary administrative units (SAU). In the same way, there may be third level units (TAU) and so forth.
We give three examples of occurrence of the Lavalette distribution: the Nigeria (NRG) population  of
local \-go\-vern\-ment areas (SAU) and the municipality population (TAU), below province and
autonomous community, within the Spanish provinces of Madrid and C\'{a}diz. We chose 
these examples after analysing population data from many countries in the world 
and picking those that are best fitted by the Lavalette function. We emphasize that 
we do not claim the Lavalette distribution to be ubiquitous in any way; our purpose 
is to show that there are some datasets where it can be a good statistical model. 

\begin{figure}
 \epsfig{file=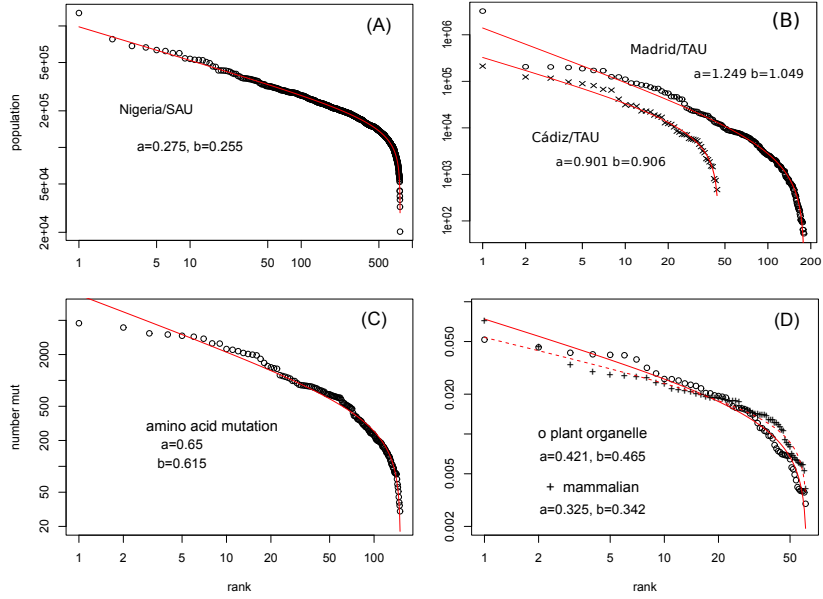,width=0.7\linewidth}
\caption{\label{fig4}
{\bf Ranked datasets fitted by Lavalette rank function.}
(A) Nigeria (NRG) local government area (the secondary administrative unit (SAU)) population;
(B) Madrid and C\'{a}diz municipality (the tertiary administrative unit (TAU)) population;
(C) Amino acid to amino acid mutation counts in the 1000 Genomes Project;
(D) Averaged codon usage (excluding the three stop codons) of plant organelles and mammals.
}
\end{figure}

Fig.\ref{fig4}(A,B) shows the rank frequency distribution of the NRG/SAU, Madrid/TAU and C\'{a}diz/TAU
population in log-log scale.  The fitted parameter values $(a,b)$ by Eq.(\ref{eq-beta}) are (0.275, 0.255) for NRG/SAU, (1.249, 1.049) for Madrid/TAU, (0.901, 0.906) for C\'{a}diz/TAU, all with
$a \approx b$. Clearly these do not follow a power law distribution. Although city population is one
of the well known examples of Zipf's law \cite{krugman,jiang}, there is a difference between cities and administrative units. The origin of Zipf's law in population and economic phenomena might be explained
by a proportionate-growth random process \cite{gabaix}. For the particular case of well separated cities, as well as firm sizes, birth and death processes explain the origin and robustness of Zipf's law \cite{saichev}. However, when regions are artificially partitioned, such as the case of administrative units, the argument for power-law may fail. Indeed, the bad fitting performance of Zipf's law on data in some counties \cite{soo,cities} might be caused by the artificial boundary in defining a city \cite{holmes}.  This leaves room for alternative functional form such as DGBD. \cite{beta-plos1}.

The second example is the amino acid mutation rates \cite{debeer} based on the amino acid changing (missense) variants in the 1000 Genomes Project \cite{1kg}. A missense mutation is a point mutation which results in the codification of a different amino acid. Because the variants are observed in normal human population with a short evolutionary history, it can be considered as an instantaneous mutation rate. The substitution rate between different species, such as the point accepted mutation (PAM) \cite{pam},
cover a much longer evolutionary history with stronger selection constraints. Out of 380 ( = 20 $\times$ 19) possible mutations between 20 amino acids, only $N=$ 150 are allowed from the single base mutation in the DNA sequence, due to the nature of the genetic code. Fig.\ref{fig4}(C) shows the ranked
amino acid to amino acid frequencies derived from the missense variants in DNA sequence of the 1000 Genomes Project. Fig.\ref{fig4}(C) shows a fitting by the Beta rank function Eq.(\ref{eq-beta}) with $a \approx 0.650$ and $b \approx 0.615$, which is again a good Lavalette function.

The third example is the codon usage of $N=$ 61  non-stop codons, with data from the Codon Usage Database \cite{codonud}. Codon usage refers to the frequency of occurrence of 181 each type of codon within a DNA sequence. We picked the two examples best demonstrating Lavalette function: genes in plant organelles (9221 species) and in (non-primate, non-rodent) mammalian nucleus (433 species). The codon frequencies are averaged over all species in plant organelle and mammalian separately. The three stop codons are discarded. The $(a,b)$ are (0.422, 0.465) for plant organelles, and (0.325,0.342) for mammalian (Fig.\ref{fig4}(D)).

With the previous examples we have illustrated the occurrence of the Lavalette
distribution. Next we propose a statistical criterion to discern if this distribution is
consistent with the data.  

\subsection*{Goodness of Fit Tests}

The first clue that a certain dataset may be well described by the Lavalette distribution
is to fit the data to DGBD function Eq.(\ref{eq-beta}), estimate the parameters and check
if $a\approx b$. If this is the case, the data set is a candidate for the Lavalette distribution.
This is a first criterion and it serves to rule out many datasets; however, it is by no means strong statistical evidence to claim the the Lavalette is a good model for the data. 

To test more rigorously whether a Lavalette function fits the observed data well, we use a re-sampling approach as discussed in \cite{clauset} which can also be called a {\sl bootstrap} \cite{efron}.
We first fit the data by the Lavalette function (Eq.(\ref{eq-lavalette})). The difference between the observed and fitted value is measured by the Kolmogorov-Smirnov (KS) distance. Using the fitted Lavalette rank function, artificial data (replicates) are generated multiple times: each time a new Lavalette rank function is fitted and KS distance calculated. The proportion of replicates with larger KS distances than the observed one is the empirical $p$-value.

A large empirical $p$-value indicates that there is not enough evidence to reject the Lavalette function.
Empirical $p$-values from 1000 replicates are 0.49 for NRG/SAU, 0.91 for Madrid/TAU, 0.88 for C\'{a}diz/TAU, 0.06 for mutation rate,  and 0.4 for codon usage in both plant organelle and mammals. These values depend on many specific choices used, e.g. how to handle replicates which have the same KS distance as the observed one, using KS distance instead of some other measure of difference between two curves, the
number of replicates, etc. The empirical $p$-value we have indicate that Eq.(\ref{eq-lavalette}) is a good fitting function for these data. 

There have been debates in the literature whether Zipf's law results 
from the central limit theorem \cite{perline,troll,mitzenmacher}.
Given a dataset, the best answer to that debate is to pick the better
fitting model between power law and lognormal distribution \cite{male}.
The approximate equivalence between Lavalette distribution and lognormal
distribution provides us with a simple method in deciding if a set of data follows
Zipf's law or lognormal distribution. For the fitting of ranked data by
the Beta rank function Eq.(\ref{eq-beta}), if $b \approx 0$, the Zipf's
law is better; if $a \approx b >0$, lognormal is better;
and if $ a \ne b \gg 0 $, neither are good fitting functions.

For our examples to illustrate the Lavalette distribution in
real data, it is obvious that lognormal distribution is a better
fitting function than the Zipf's law. We can further  quantify the fitting
performance by model selection techniques such as Akaike information criterion
(AIC) \cite{aic_book,aic,wli_prl}, with the better model exhibiting lower AIC
value.  The AIC$_{lav}$-AIC$_{zipf}$= $ N \log(SSE_{lav}/SSE_{zipf})$ \cite{beta-jql1},
where SSE is the sum squared error, is -3284.6,-410.3, -108.1  for the NRG/SAU, 
 Madrid/TAU, C\'{a}diz/TAU data, -353.7 for the amino acid mutate data, and
-101.7, -114.7 for plant organelle, mammalian codon frequencies,
all representing an overwhelming support to the Lavalette function
or lognormal distribution over the Zipf's law.

\section*{Discussion and Conclusions}
\label{sec:conclusions}

We have presented a novel probability distribution function and showed that it is a good 
alternative for data that does not follow a perfect Zipf's law. We have seen that this distribution
yields a very good approximation to the lognormal distribution. 
Although it is perhaps less important because of the approximate equivalence between
the Lavalette and the lognormal distributions, one may still sometimes want to determine whether a
data is better fitted by the Lavalette or lognormal distribution.
We propose the following procedure for this test: (i) log-transform, then
standardize (zero mean, unit standard deviation) the raw data, to $x'$;
(ii) compare the empirical cdf of $x'$ to both standard normal
and logistic distribution cdf with a scaling parameter $\alpha = \pi/\sqrt{3}$
($cdf=1/(1+e^{-\alpha x' })$); (iii) if the standard normal function is closer
to the data, lognormal distribution fits the original data better, otherwise,
Lavalette function is better. Using this procedure and KS distance as the
measure of difference, NRG/SAU is the only data which Lavalette is better
than lognormal distribution. If sum of absolute error is used to measure
the difference,  codon usage of mammals is another data which prefers Lavalette over lognormal.

In conclusion, by connecting Lavalette function to lognormal distribution,
we achieve a better understanding of the DGBD function and the limitations of the Zipf's law.


\section*{Materials and Methods}

Population data for administrative units and sub-units in a large sample of countries is
available in the database Statoids \url{http://www.statoids.com} (accessed April 2016).
Population of local government areas (SAU) in Nigeria were taken from this database. Spain's population
at PAU, SAU and TAU levels are available from the National Statistics Institute (INE), \url{http://www.ine.es/en/pob_xls/pobmun12_en.xls}. We chose these examples after analysing population data from many countries in the world and picking those that are best fitted by the Lavalette function.

Amino acid to amino acid mutation rates were calculated from missense variants, taken from DNA sequence of the 1000 genomes project, available at \url{http://www.1000genomes.org/}. In particular, we used mutation data from \url{http://journals.plos.org/ploscompbiol/article?id=10.1371/journal.pcbi.1003382}
fig (1). From this data, we counted the relative frequency of occurrence for each mutation.

We calculated codon usage of 61 non-stop codons for genes in plant organelles and
non-primate and non-rodent mammalian nucleus. Data were downloaded from Codon
Usage Database \url{http://www.kazusa.or.jp/codon/}.

All the data used in our analysis is available on \url{https://figshare.com/articles/Data_rar/3363961}.

\section*{Acknowledgements}

This project was partially supported by PAPIIT/UNAM IN107414. OF acknowledges
financial support from CONACyT Mexico and is grateful to Manuel Falconi for helpful
discussion and constructive comments. PM wishes 
to thank the PASPA/UNAM program.

\section*{Author Contributions}

Conceived and designed the experiments OF PM GC WL. Performed the experiments
OF WL. Analysed the data OF YY WL. Wrote the paper OF WL.

\bibliography{references}

\end{document}